\documentclass[nohyper]{JHEP3}

\usepackage{amsmath,amssymb}
\usepackage{graphicx}
\usepackage{cite}

\newcommand{\be}{\begin{equation}}
\newcommand{\ee}{\end{equation}}
\newcommand{\bea}{\begin{eqnarray}}
\newcommand{\eea}{\end{eqnarray}}
\newcommand{\epk}{\epsilon_{k}}

 \newcommand{\fh}{f_h^{eq}}
 \newcommand{\fht}{f_{\tilde{h}}^{eq}}
\newcommand{\fLeq}{f_{\ell}^{eq}} 
\newcommand{\fLteq}{f_{\widetilde{\ell}}^{eq}}

\title{On Gaugino Contributions to Soft Leptogenesis}
\author{Chee Sheng Fong 
\\
C.N. Yang Institute for Theoretical Physics\\
  State University of New York at Stony Brook\\
  Stony Brook, NY 11794-3840, USA,\\
  E-mail: \email{fong@insti.physics.sunysb.edu}}
\author{M.~C.~Gonzalez-Garcia\\
  Instituci\'o Catalana de Recerca i Estudis Avan\c{c}ats (ICREA), \\
  Departament d'Estructura i Constituents de la Mat\`eria,
  Universitat de Barcelona,\\
  Diagonal 647, E-08028 Barcelona, Spain\\
{\rm and:}  \\
C.N. Yang Institute for Theoretical Physics\\
  State University of New York at Stony Brook\\
  Stony Brook, NY 11794-3840, USA,\\
  E-mail: \email{concha@insti.physics.sunysb.edu}}

\abstract{
We study the contributions to CP violation in right-handed sneutrino decays 
induced by soft supersymmetry-breaking gaugino masses including
flavour effects and paying special attention to the role of thermal 
corrections. 
Using a field-theoretical as well as a quantum mechanical 
approach we compute the CP asymmetries and we 
conclude that for all the soft-supersymmetry 
breaking sources of CP violation considered,    
an exact cancellation between the asymmetries produced 
in the fermionic and bosonic channels occurs 
at  $T=0$ up to second order in 
soft supersymmetry-breaking parameters.   
Once thermal effects are included the new sources of
CP violation induced by supersymmetry-breaking gaugino masses
can be sizeable and they can produce the observed baryon asymmetry
for conventional values of the $B$ parameter.}
\preprint{%
  YITP-SB-08-48\\}

\begin{document}

\section{Introduction}
The discovery of neutrino oscillations makes leptogenesis a very
attractive solution to the baryon asymmetry problem \cite{fy,leptoreview}.  
In the {\sl standard} type I seesaw  framework \cite{ss}, the   
singlet heavy neutrinos have  lepton number violating 
Majorana masses and when decay out of equilibrium produce 
dynamically a lepton asymmetry which  is partially converted into a
baryon asymmetry due to fast sphaleron processes.

For a hierarchical spectrum of right-handed neutrinos,  
successful leptogenesis requires generically quite heavy singlet neutrino 
masses~\cite{di}, of order $M>2.4 (0.4)\times 10^9$~GeV for vanishing
(thermal) initial neutrino densities~\cite{di,Mbound} 
(although flavour effects \cite{flavour1,flavour2,db2,oscar} 
and/or extended scenarios \cite{db1,ma} may affect this limit).
Low-energy supersymmetry can be invoked to naturally 
stabilize  the hierarchy between this new scale and the 
electroweak one. This, however, introduces a certain conflict between 
the gravitino bound on the reheat temperature and the thermal 
production of right-handed neutrinos \cite{gravi}. 
A way out of this conflict is provided by resonant leptogenesis~\cite{PU}. 
In this scenario right-handed neutrinos are nearly degenerate in mass 
which makes the 
self energy contributions to the  CP asymmetries resonantly enhanced 
and allowing leptogenesis to be possible at much lower temperatures.

Once supersymmetry has been introduced, leptogenesis is induced also
in singlet sneutrino decays.  If supersymmetry is not broken, the
order of magnitude of the asymmetry and the basic mechanism are the
same as in the non-supersymmetric case. However,  
as shown in  Refs.\cite{soft1,soft2}, supersymmetry-breaking 
terms can induce effects which are essentially different 
from the neutrino ones.  In brief, soft supersymmetry-breaking terms
involving the singlet sneutrinos remove the mass degeneracy between
the two real sneutrino states of a single neutrino generation, and
provide new sources of lepton number and CP violation. 
In this case, as for the case of resonant 
leptogenesis, it is the  sneutrino self-energy contributions 
to the  CP asymmetries which are resonantly enhanced. 
As a consequence, the mixing between the sneutrino states can generate a
sizable CP asymmetry in their decays. 
This scenario was  termed ``soft leptogenesis''.
Altogether it was found that the asymmetry is large for a right-handed
neutrino mass scale relatively low, in the range $10^{5}-10^{8}$ GeV,
well below the reheat temperature limits, what solves the cosmological
gravitino problem.  However in order to generate
enough asymmetry the lepton-violating  soft bilinear coupling,
$B$, responsible for the sneutrino mass splitting, has to be
unconventionally small~\cite{soft1,soft2,oursoft,ourqbe}
\footnote{Extended scenarios~\cite{ourinvsoft,softothers} may alleviate the
unconventionally-small-$B$ problem.}.

In soft leptogenesis induced by CP violation in mixing
as discussed above an exact cancellation occurs between the asymmetry 
produced in the fermionic and bosonic channels at $T=0$. 
Thermal effects, thus, play a fundamental role in this mechanism:   
final-state Fermi blocking and Bose stimulation as well as effective masses 
for the particle excitations in the plasma 
break supersymmetry and effectively remove this degeneracy. 

In Ref.~\cite{soft3} the possibility of soft leptogenesis 
generated by CP violation in  right-handed sneutrino decay and in 
the interference of mixing and decay was considered. 
These new sources of CP violation (the so called  ``new ways 
to soft leptogenesis'') are induced by vertex
corrections due to gaugino soft supersymmetry-breaking masses.  
Some of these contributions, although suppressed by a loop factor and higher
order in the supersymmetry-breaking parameters are relevant because they 
can be sizeable for natural values of the  $B$ parameter. 
Furthermore it was found that,
unlike for CP violation in mixing, these contributions did not 
require thermal effects as they did not vanish at $T=0$.

In this work we revisit the role of thermal effects in soft leptogenesis 
due to CP violation in  right-handed sneutrino decays induced by gaugino soft 
supersymmetry-breaking masses.  In Sec.\ref{sec:QFT}
we describe the one-generation see-saw model in the presence of the soft 
supersymmetry-breaking terms and compute the relevant CP asymmetries
in a field-theoretical approach. We find that for all soft 
supersymmetry-breaking 
sources of CP violation considered, at  $T=0$ the exact cancellation 
between the asymmetries produced in the fermionic and bosonic channels holds 
up to second order in  soft supersymmetry-breaking parameters.   
In Sec.\ref{sec:QM} we recompute the asymmetries using a quantum mechanical
approach, based on an effective (non hermitic) Hamiltonian. 
We find  the same $T$ dependence of the resulting CP asymmetries. 
Finally in Sec.~\ref{sec:results} we present our
quantitative results and determine  the region of parameters in which
successful leptogenesis induced by the different contributions to the
CP asymmetry is possible including the dominant thermal corrections
as well as flavour-dependent effects associated  
with the charged lepton Yukawa couplings in this scenario. 

\section{The CP Asymmetry: Field Theoretical Approach}
\label{sec:QFT}

The supersymmetric see-saw model could be described by the superpotential:
\begin{equation}
W=\frac{1}{2}M_{ij}N_{i}N_{j}+Y_{ij}
\epsilon_{\alpha\beta}N_{i}L_{j}^{\alpha}H^{\beta},
\label{eq:superpotential}
\end{equation}
where $i,j=1,2,3$ are flavour indices and $N_{i}$, $L_{i}$, $H$
are the chiral superfields for the right-handed (RH) neutrinos, 
the left-handed (LH)
lepton doublets and the Higgs doublets with 
$\epsilon_{\alpha\beta}=-\epsilon_{\beta\alpha}$
and $\epsilon_{12}=+1$. 

The relevant soft breaking terms involving the RH sneutrinos
$\widetilde{N_{i}}$ and SU(2) gauginos $\tilde{\lambda}_{2}^{a}$ are
given by \footnote{The effect of
$U(1)$ gauginos can be included in similar form.} 
\begin{eqnarray} \mathcal{L}_{soft} & = 
&
-\left(A_{ij}Y_{ij}\epsilon_{\alpha\beta}\widetilde{N}_{i}
\tilde{\ell}_{j}^{\alpha}h^{\beta}+\frac{1}{2}B_{ij}
M_{ij}\widetilde{N}_{i}\widetilde{N}_{j}
+\frac{1}{2}m_{2}\overline{\tilde{\lambda}}_{2}^{a}
P_{L}\tilde{\lambda}_{2}^{a}
+\mbox{h.c.}\right)\; .
\label{eq:soft_terms}\end{eqnarray}

The Lagrangian for interaction terms involving RH sneutrinos 
$\widetilde{N}_{i}$, the RH neutrinos $N_{i}$ 
and the $\tilde{\lambda}_{2}$ with (s)leptons
and higgs(inos) can be written as:
\begin{eqnarray}
\mathcal{L}_{int}  =  -Y_{ij}\epsilon_{\alpha\beta}
&&\left( 
M_{i}\widetilde{N}_{i}^{*}\tilde{\ell}_{j}^{\alpha}h^{\beta}
+\overline{\tilde{h}}^{\beta}P_{L}\ell_{j}^{\alpha}\widetilde{N}_{i}
+\overline{\tilde{h}}^{\beta}P_{L}N_{i}\tilde{\ell}_{j}^{\alpha}
+\overline{N}_i P_L\ell_j^{\alpha}h^{\beta}
+A \widetilde{N}_{i}\tilde{\ell}_{j}^{\alpha}h^{\beta}\right)
 \nonumber \\ 
 -g_{2} && \left(
\overline{\tilde{\lambda}}_{2}^{\pm}P_{L}
(\sigma_{1})_{\alpha\beta}\ell_{i}^{\alpha}\tilde{\ell}_{i}^{\beta*}
-\frac{1}{\sqrt{2}}\overline{\tilde{\lambda}}_{2}^{0}P_{L}
(\sigma_{3})_{\alpha\beta}\ell_{i}^{\alpha}\tilde{\ell}_{i}^{\beta*}
 \right. \nonumber \\
 &  & \left.
\;\;\;\overline{\tilde{h}}^{\alpha}P_{L}(\sigma_{1})_{\alpha\beta}
\tilde{\lambda}_{2}^{\pm}h^{\beta*}-\frac{1}{\sqrt{2}}
\overline{\tilde{h}}^{\alpha}P_{L}(\sigma_{3})_{\alpha\beta}
\tilde{\lambda}_{2}^{0}h^{\beta*}\right)
+\mbox{h.c.}.
\label{eq:int_basis}
\end{eqnarray}
$\ell_{i}^{T}=\left(\nu_{i},\ell_{i}^{-}\right)$, 
$\tilde{\ell}_{i}^{T}=\left(\tilde{\nu}_{i},\tilde{\ell}_{i}^{-}\right)$
are the  lepton and slepton doublets and, 
$h^{T}=\left(h^{+},h^{0}\right)$ and
$\tilde{h}^{T}=\left(\tilde{h}^{-},\tilde{h}^{0}\right)$,
are the  Higgs and higgsino doublets.   
$\tilde{\lambda}_{2}^{\pm}$ denotes $\tilde{\lambda}_{2}^{+}$
for $\alpha\beta=01$ and $\tilde{\lambda}_{2}^{-}$ for $\alpha\beta=10$
with $\sigma_{1,3}$ being the Pauli matrices, 
and $P_{L,R}$ are  the left or right projection operator.

The sneutrino and antisneutrino states mix with mass eigenvectors
\begin{eqnarray}
\widetilde{N}_{+i} & = &
\frac{1}{\sqrt{2}}(e^{i\Phi/2}\widetilde{N}_{i}+e^{-i\Phi/2}
\widetilde{N}_{i}^{*}),\nonumber
\\ \widetilde{N}_{-i} & = &
\frac{-i}{\sqrt{2}}(e^{i\Phi/2}
\widetilde{N}_{i}-e^{-i\Phi/2}\widetilde{N}_{i}^{*}),
\label{eq:mass_eigenstates}
\end{eqnarray}
where $\Phi\equiv\arg(BM)$ and with mass eigenvalues
\begin{eqnarray}
M_{ii\pm}^{2} & = & M_{ii}^{2} \pm|B_{ii}M_{ii}|.
\label{eq:mass_eigenvalues}
\end{eqnarray}

From \eqref{eq:int_basis} and \eqref{eq:mass_eigenstates}, we can
write down the Lagrangian in the mass basis as
\begin{eqnarray}
\mathcal{L}_{int}=
-Y_{ij}\epsilon_{\alpha\beta}&& \left\{\frac{1}{\sqrt{2}} 
\widetilde{N}_{+i}\left[\overline{\tilde{h}}^{\beta}P_{L}\ell_{j}^{\alpha}
+(A_{ij}+M_{i})\tilde{\ell}_{j}^{\alpha}h^{\beta}\right] \right.
\nonumber \\
&& \left.
+\frac{i}{\sqrt{2}}\widetilde{N}_{-i}\left[\overline{\tilde{h}}^{\beta}P_{L}\ell_{_{j}}^{\alpha}
+(A_{ij}-M_{i})\tilde{\ell}_{j}^{\alpha}h^{\beta}\right]
+\overline{\tilde{h}}^{\beta}P_{L}N_{i}\tilde{\ell}_{j}^{\alpha}
+\overline{N}_i P_L\ell_j^{\alpha}h^{\beta}\right\}
\nonumber \\
&& -g_{2}\left(
\overline{\tilde{\lambda}}_{2}^{\pm}P_{L}
(\sigma_{1})_{\alpha\beta}\ell_{i}^{\alpha}\tilde{\ell}_{i}^{\beta*}
-\frac{1}{\sqrt{2}}\overline{\tilde{\lambda}}_{2}^{0}P_{L}
(\sigma_{3})_{\alpha\beta}\ell_{i}^{\alpha}\tilde{\ell}_{i}^{\beta*}
\right.\nonumber \\
 &  & \left.
+\overline{\tilde{h}}^{\alpha}P_{L}(\sigma_{1})_{\alpha\beta}
\tilde{\lambda}_{2}^{\pm}h^{\beta*}-\frac{g_{2}}{\sqrt{2}}
\overline{\tilde{h}}^{\alpha}P_{L}(\sigma_{3})_{\alpha\beta}
\tilde{\lambda}_{2}^{0}h^{\beta*}\right)
+\mbox{h.c.}.
\label{eq:mass_basis}
\end{eqnarray}

In what follows, we will consider a single generation of $N$ 
and $\widetilde{N}$ which we label as $1$. 
We also assume proportionality of soft trilinear terms and drop the
flavour indices for the coefficients $A$ and $B$.  

As discussed in Refs.~\cite{soft1,soft2,soft3}, in this case,  
after superfield 
rotations the Lagrangians (\ref{eq:superpotential}) and (\ref{eq:soft_terms}) 
have two independent physical CP violating phases:
\begin{eqnarray}
\phi_A={\rm arg}(A B^*), \\
\phi_{g}=\frac{1}{2}{\rm arg}(B m_2^*), 
\label{eq:CPphase}
\end{eqnarray}
which we choose to assign to $A$ and to the gaugino coupling 
operators  (the last two lines which are multiplied by $g_2$ 
in Eq.\eqref{eq:mass_basis} ) respectively. So for the calculations
below  we will take $M$, $B$, $m_2$ and $Y_{1k}$ to be positive real and
$A$ with phase $\phi_A$ and define a complex coupling 
$\tilde g_2=g_2\exp(i\phi_g)$ respectively.

As discussed in Ref.\cite{soft2}, when $\Gamma \gg \Delta
M_{\pm}\equiv M_+ -M_- $, the two singlet sneutrino states are not
well-separated particles. In this case, the result for the asymmetry
depends on how the initial state is prepared. In what follows we will
assume that the sneutrinos are in a thermal bath with a thermalization
time $\Gamma^{-1}$ shorter than the typical oscillation times, $\Delta
M_\pm^{-1}$, therefore coherence is lost and it is appropriate to
compute the CP asymmetry in terms of the mass eigenstates
 Eq.(\ref{eq:mass_eigenstates}).

\begin{figure}
\begin{center}
\includegraphics[scale=0.7]{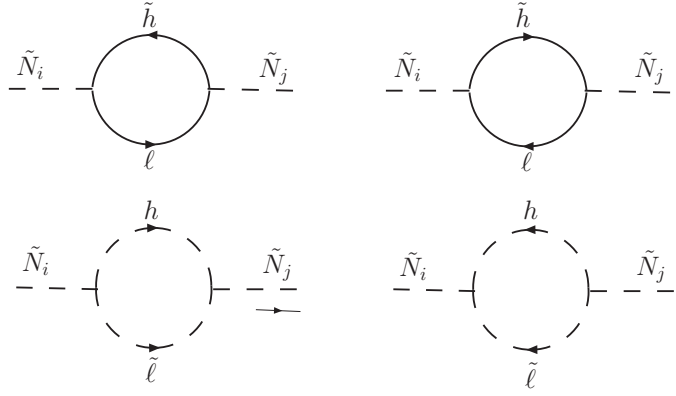}
\end{center}
\caption{Feynman diagrams contributing to the RH sneutrino self-energies
at one-loop.}
\label{fig:self_energies}
\end{figure}

\begin{figure}
\begin{center}
\includegraphics[scale=0.7]{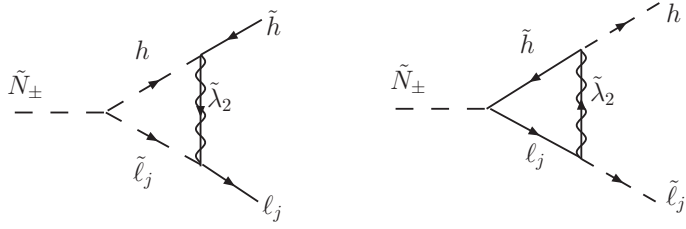}
\caption{Feynman diagrams contributing to the RH sneutrino decay vertex 
at one-loop.}
\end{center}
\label{fig:vertex_gauginos}
\end{figure}

We compute the relevant decay amplitudes following the effective field-theoretical
approach described in \cite{pi}, which takes into account the CP
violation due to mixing and decay (as well as their interference) 
of nearly degenerate states by using resummed
propagators for unstable mass eigenstate particles.  The decay
amplitude $\hat{\mathcal A}_i^{a_k}$ of the unstable external state
$\widetilde{N}_i$ defined in Eq.~(\ref{eq:mass_eigenstates}) into a final
state $a_k$ ($a_k\equiv s_k,f_k$ with 
$s_k=\tilde{\ell}^\alpha_k h^\beta$ and 
$f_k=\ell_k^\alpha \tilde h^\beta$)
is described by a superposition of amplitudes with stable
final states: 
\begin{eqnarray}
\hat{\mathcal{A}}_{\pm}^{a_{k}} & = 
& \left(A_{\pm}^{a_k}+i{\mathcal{V}_{\pm}^{a_k}}^{\mbox{abs}}
(M^2_\pm)
\right)
-\left(A_{\mp}^{a_k}+i{\mathcal{V}_{\mp}^{a_k}}^{\mbox{abs}}
(M^2_\pm)\right)
\frac{i\Sigma_{\mp\pm}^{\mbox{abs}}}{M_\pm^{2}-M_{\mp}^{2}
+i\Sigma_{\mp\mp}^{\mbox{abs}}},\label{eq:amp}\\
\overline{\hat{\mathcal{A}}_{\pm}^{\bar{a}_{k}}} 
& = & \left({A_{\pm}^{a_k}}^{*}+i{\mathcal{V}_{\pm}^{a_k}}^{\mbox{abs}*}
(M^2_\pm)\right)
-\left({A_{\mp}^{a_k}}^{*}+i{\mathcal{V}_{\mp}^{a_k}}^{\mbox{abs}*}
(M^2_\pm)\right)
\frac{i\overline{\Sigma}_{\mp\pm}^{\mbox{abs}}}{M_\pm^{2}-M_{\mp}^{2}
+i\overline{\Sigma}_{\mp\mp}^{\mbox{abs}}}\;.
\label{eq:amp_cp} 
\end{eqnarray}
$A_{\pm}^{a_k}$ are the tree-level amplitudes:
\begin{eqnarray}
&& A_{+}^{s_k} =  \frac{Y_{1k}}{\sqrt{2}}(A^{*}+M)\epsilon_{\alpha\beta},
\;\;\;\;\;\;\;\;\;\;\;\;\;\;\;\;\;\;\;\;\;\;\;
A_{-}^{s_k} =  -i\frac{Y_{1k}}{\sqrt{2}}(A^{*}-M)\epsilon_{\alpha\beta}, \\
&& A_{+}^{f_k} =  
\frac{Y_{1k}}{\sqrt{2}}[\bar{u}(p_{\ell})
P_{R}v(p_{h})] \epsilon_{\alpha\beta},
\;\;\;\;\;\;\;\;\;\;\;\;\;\;\;\;\;
A_{-}^{f_k} =  -i\frac{Y_{1k}}{\sqrt{2}}
[\bar{u}(p_{\ell})P_{R}v(p_{h})]\epsilon_{\alpha\beta}.
\end{eqnarray}
$\Sigma_{ab}^{\mbox{abs}}$ are the absorptive parts of 
the $\widetilde{N}_{b}\to\widetilde{N}_{a}$ self-energies  
(see Fig.~\ref{fig:self_energies}):
\begin{eqnarray}
\Sigma_{\mp\mp}^{(1)\mbox{abs}}& = & 
\Gamma M\left[\frac{1}{2}+\frac{M_\pm^{2}}{2M^{2}}+\frac{|A|^{2}}{2M^{2}}
\mp\frac{\mbox{Re}(A)}{M}\right], \\
\Sigma_{\mp\pm}^{(1)\mbox{abs}} & = & 
-\Gamma\mbox{Im}(A),
\label{eq:self_energies}
\end{eqnarray}
with
\be
\Gamma =\frac {\displaystyle \sum_k M |Y_{1k}|^2}{\displaystyle 4 \pi}
= \frac {\displaystyle M (YY^\dagger)_{11}}{\displaystyle 4 \pi}  .
\label{eq:gamma}
\ee
${\mathcal{V}_{\pm}^{a_k}}^{\mbox{abs}}$ are the absorptive
parts of the vertex corrections (see Fig.~\ref{fig:vertex_gauginos}):
\begin{eqnarray}
{\mathcal{V}_{+}^{s_k}}^{\mbox{abs}}\left(p^2\right)& 
= & \frac{Y_{1k}}{\sqrt{2}}\frac{3m_2}{32\pi}(\tilde g_{2})^{2}
\ln\frac{m_{2}^{2}}{p^{2}+m_2^{2}}
\epsilon_{\alpha\beta}\; , \\
{\mathcal{V}_{-}^{s_k}}^{\mbox{abs}}\left(p^2\right)& = & -i\frac{Y_{1k}}{\sqrt{2}}
\frac{3 m_{2}}{32\pi}(\tilde g_{2})^{2}
\ln\frac{m_{2}^{2}}{p^{2}+m_2^{2}}
\epsilon_{\alpha\beta}\; , \\
{\mathcal{V}_{+}^{f_k}}^{\mbox{abs}}\left(p^2\right)& = & 
\frac{Y_{1k}}{\sqrt{2}}\frac{3 m_2}{32\pi p^{2}}(A^{*}+M)
(\tilde g_{2}^{*})^{2}\ln\frac{m_2^{2}}{p^{2}+m_2^{2}}
[\bar{u}(p_{\ell})P_{R}v(p_{h})]\epsilon_{\alpha\beta} 
,\\
{\mathcal{V}_{-}^{f_k}}^{\mbox{abs}}\left(p^2\right)& = & -i\frac{Y_{1k}}{\sqrt{2}}
\frac{3 m_2}{32\pi p^{2}}(A^{*}-M)(\tilde g_{2}^{*})^{2}
\ln\frac{m_2^{2}}{p^{2}+m_2^{2}}
[\bar{u}(p_{\ell})P_{R}v(p_{h})]\epsilon_{\alpha\beta} \; .
\label{eq:vertexabs}
\end{eqnarray}
\footnote{We notice that there is an irrelevant global $i$  
factor in the tree level
$A_{-}^{a_k}$ and one-loop ${\mathcal{V}_{-}^{a_k}}^{\mbox{abs}}$
amplitudes  compared to 
$A_{+}^{a_k}$ and ${\mathcal{V}_{+}^{a_k}}^{\mbox{abs}}$
arising from the particular choice of 
global phase in the definition of $\widetilde N_-$ in 
Eq.\eqref{eq:mass_eigenstates}.}.  

\subsection{The CP asymmetry}

The CP asymmetry produced in the decay of the 
states $\widetilde{N}_{i=\pm}$ which enters into the Boltzmann Equations (BE) is given by:
\be
\epk = \frac{\displaystyle \sum_{i=\pm,a_k} 
\gamma(\widetilde{N}_i \rightarrow a_k)
- \gamma(\widetilde{N}_i \rightarrow \bar{a}_k)}
{\displaystyle \sum_{i=\pm,a_k,k} \gamma(\widetilde{N}_i \rightarrow a_k)
+ \gamma(\widetilde{N}_i \rightarrow \bar{a}_k)} \ , 
\label{eq:cp_asym_total} 
\ee
where we denote by $\gamma$ the thermal averaged rates.  
In the rest frame of $\widetilde{N}_{\pm}$, \eqref{eq:cp_asym_total} 
simplifies to:
\begin{equation}
\epk=\epsilon_{+ k}^{s}+\epsilon_{- k}^{s}+\epsilon_{+k}^{f}+\epsilon_{-k}^{f},
\end{equation}
where 
\begin{eqnarray} 
\epsilon^{s}_{\pm k} =\frac{\displaystyle
\left(|\hat{\mathcal{A}}_{\pm}^{s_{k}}|^{2}
-|\overline{\hat{\mathcal{A}}_{\pm}^{\bar{s}_{k}}}|^{2}\right)
c^{s_{k}}_\pm/M_\pm}{\displaystyle \sum_{i=\pm,a_{k},k}
\left(|\hat{\mathcal{A}}_i^{a_{k}}|^{2}
+|\overline{\hat{\mathcal{A}}_{i}^{\bar{a}_{k}}}|^{2}\right)
c^{a_{k}}_i/M_{i}}
\;,
\label{eq:asyms} 
\\
\epsilon_{\pm k}^{f} 
=\frac{\left(|\hat{\mathcal{A}}_{\pm}^{f_{k}}|^{2}
-|\overline{\hat{\mathcal{A}}_{\pm}^{\bar{f}_{k}}}|^{2}\right)
c^{f_{k}}_\pm/M_\pm}{\displaystyle \sum_{i=\pm,a_{k},k}
\left(|\hat{\mathcal{A}}_{i}^{a_{k}}|^{2}
+|\overline{\hat{\mathcal{A}}_{i}^{\bar{a}_{k}}}|^{2}\right)
c^{a_{k}}_i/M_{i}} \;. 
\label{eq:asymf} 
\end{eqnarray}
In Eqs.(\ref{eq:asyms}) and (\ref{eq:asymf})
$c^{s_k}_i, c^{f_k}_i$ are the phase-space factors of the scalar and 
fermionic channels, respectively.  
As long as we neglect the zero temperature lepton and slepton masses
and  small Yukawa couplings,  
these phase-space factors are flavour independent
and they are the same for $i=\pm$. 
After  including finite temperature effects in the approximation of  
decay at rest of the $\widetilde N_\pm$ they are given by:
\bea
c^f_+(T)=c^f_-(T)\equiv c^f(T)
&=&(1-x_{\ell} -x_{\tilde{h}})\lambda(1,x_{\ell},x_{\tilde{h}})
\left[ 1-\fLeq\right] \left[ 1-\fht\right], 
\label{cfeq}\\
c^s_+(T)=c^s_-(T)\equiv c^s(T)&=&\lambda(1,x_h,x_{\tilde{\ell}})
\left[ 1+\fh\right] \left[ 1+\fLteq\right],
\label{cbeq}
\eea
where
\bea
f^{eq}_{h,\tilde{\ell}}&=&\frac{1}{\exp[E_{h,\tilde{\ell}}/T]-1},
\label{eq:fHeq}\\
f^{eq}_{\tilde h,\ell}&=& \frac{1}{\exp[E_{\tilde h,\ell}/T]+1},  
\label{eq:fheq}
\eea
are the  Boltzmann-Einstein and Fermi-Dirac equilibrium distributions,
respectively, and 
\bea
&E_{\ell,\tilde h}=\frac{M}{2} (1+x_{\ell,\tilde{h}}-
x_{\tilde h,\ell}), ~~~
E_{h,\tilde{\ell}}=\frac{M}{2} (1+x_{h ,\tilde{\ell}}-
x_{\tilde{\ell},h}),&\\
&\lambda(1,x,y)=\sqrt{(1+x-y)^2-4x},~~~
x_a\equiv \frac{m_a(T)^2}{M^2}.&
\eea
The thermal masses for the relevant supersymmetric degrees of
freedom are \cite{thermal}:
\bea
m_h^2(T)=2 m_{\tilde h}^2(T)&=& \left(\frac{3}{8}g_2^2+\frac{1}{8}g_Y^2
+\frac{3}{4}\lambda_t^2\right) \, T^2\; ,\\
m_{\tilde{\ell}}^2(T)=2 m_\ell^2(T)&=& \left(\frac{3}{8}g_2^2+\frac{1}{8}g_Y^2
\right)\, T^2\; .
\eea
Here $g_2$ and $g_Y$ are gauge couplings 
and $\lambda_t$ is the top Yukawa,
renormalized at the appropriate high-energy scale. 

Substituting \eqref{eq:amp} and \eqref{eq:amp_cp} 
into 
\eqref{eq:asyms} and \eqref{eq:asymf} 
we get in the numerators: 
\begin{eqnarray}
|\hat{\mathcal{A}}_{\pm}^{a_{k}}|^{2}-
|\overline{\hat{\mathcal{A}}_{\pm}^{\bar{a}_{k}}}|^{2} 
&\simeq&
-4\left\{-\mbox{Im}\left[{A_{\pm}^{a_k}}^{*}A_{\mp}^{a_k}
\Sigma_{\mp\pm}^{\mbox{abs}}\right]
\frac{M_\pm^{2}-M_{\mp}^{2}}
{(M_\pm^{2}-M_{\mp}^{2})^{2}+|\Sigma_{\mp\mp}^{\mbox{abs}}|^{2}}
\right.\nonumber \\ 
 &   &
+\mbox{Im}\left[{A_{\pm}^{a_k}}^{*}{\mathcal{V}_{\pm}^{a_k}}^{\mbox{abs}}
(M^2_\pm)
\right]
\label{eq:cp_asym_num} \\
 &  & \left.+\mbox{Im}\left[{\mathcal{V}_{\pm}^{a_k}}^{\mbox{abs}*}
(M^2_\pm)
A_{\mp}^{a_k}
\Sigma_{\mp\pm}^{\mbox{abs}}-{A_{\pm}^{a_k}}^{*}
{\mathcal{V}_{\mp}^{a_k}}^{\mbox{abs}}(M^2_\pm)
\Sigma_{\mp\pm}^{\mbox{abs}}\right]
\frac{\Sigma_{\mp\mp}^{\mbox{abs}}}
{(M_\pm^{2}-M_{\mp}^{2})^{2}+|\Sigma_{\mp\mp}^{\mbox{abs}}|^{2}}\right\},
\nonumber 
\end{eqnarray}
where we have used the relations 
$\Sigma_{\mp\mp}^{\mbox{abs}}=\overline{\Sigma}_{\mp\mp}^{\mbox{abs}}$
and $\Sigma_{\mp\pm}^{\mbox{abs}*}=\overline{\Sigma}_{\mp\pm}^{\mbox{abs}}$.
The $\simeq$ sign means that terms of order $\delta_{S}^{3}$
and higher are ignored with
\begin{eqnarray}
\delta_{S} & \equiv & \frac{|A|}{M},\frac{B}{M},\frac{m_2}{M}.
\end{eqnarray}
The three lines in \eqref{eq:cp_asym_num} correspond 
respectively to  (i) CP violation 
in $\widetilde N$  mixing from the off-diagonal one-loop self-energies 
(this corresponds to the effects originally considered in 
Refs.~\cite{soft1,soft2}), 
(ii)  CP violation due to the gaugino-mediated one-loop vertex 
corrections to the $\widetilde N$ decay, and (iii) CP violation 
in the interference of  vertex and self-energies.

In the denominator of \eqref{eq:asyms} and \eqref{eq:asymf}  
we consider only the tree-level amplitudes
$|\hat{\mathcal{A}}_{\pm}^{a_{k}}|^{2}
+|\overline{\hat{\mathcal{A}}_{\pm}^{\bar{a}_{k}}}|^{2} 
= 2|A_{\pm}^{a_k}|^{2}$, with  
$|A_{\pm}^{s_k}|^{2} =  Y_{1k}^{2}
\left[|A|^{2}+M^{2}\pm 2M\mbox{Re}(A)\right]$ and $ 
|A_{\pm}^{f_k}|^{2}  =  Y_{1k}^{2}M_{\pm}^{2}$.

Using the explicit forms in Eqs.\eqref{eq:self_energies} and
\eqref{eq:vertexabs} we find that up to order $\delta_{s}^{2}$, the 
three contributions to the CP asymmetry from 
scalar and fermion decays verify:
\begin{eqnarray}
\epsilon_{\pm k}^{sS} 
=\frac{c^{s}(T)}{c^{s}(T)+c^{f}(T)} \epsilon^S_{\pm k}, 
\;\;\;\; \;\;\;\; 
\epsilon_{\pm k}^{fS} 
=-\frac{c^{f}(T)}{c^{s}(T)+c^{f}(T)} \epsilon^S_{\pm k},
\nonumber  \\
\epsilon_{\pm k}^{sV} 
=\frac{c^{s}(T)}{c^{s}(T)+c^{f}(T)} \epsilon^V_{\pm k}, 
\;\;\;\; \;\;\;\; 
\epsilon_{\pm k}^{fV} 
=-\frac{c^{f}(T)}{c^{s}(T)+c^{f}(T)} \epsilon^V_{\pm k}, 
\nonumber \\
\epsilon_{\pm k}^{sI} 
=\frac{c^{s}(T)}{c^{s}(T)+c^{f}(T)} \epsilon^I_{\pm k}, 
\;\;\;\; \;\;\;\; 
\epsilon_{\pm k}^{fI} 
=-\frac{c^{f}(T)}{c^{s}(T)+c^{f}(T)} \epsilon^I_{\pm k}, 
\label{eq:cancel}
\end{eqnarray}
with 
\begin{eqnarray}
\epsilon_{\pm k}^{S} 
& =& -K^0_{k}\frac{|A|}{M}\left(1\mp\frac{B}{2M}\right)
\sin\left(\phi_{A}\right)\frac{2B\Gamma}{4B^{2}+\Gamma^{2}},
\label{eq:cp_asym_final2}\\
\epsilon^V_{\pm k} 
&=&
-\frac{3K^0_{k}\alpha_{2}}{8}\frac{m_2}{M}
\ln\frac{m_2^2}{m_2^2+M^2}
\left[\frac{|A|}{M}
\sin\left(\phi_{A}+2\phi_{g}\right)-\frac{B}{M}
\sin\left(2\phi_{g}\right)\pm\sin\left(2\phi_{g}\right)\right],
\label{eq:cp_asym_final1}\\
\epsilon_{\pm k}^{I} 
& = & \frac{3K^0_{k}\alpha_{2}}{4}\frac{m_2}{M}\frac{|A|}{M}
\ln\frac{m_2^2}{m_2^2+M^2}\sin\left(\phi_{A}\right)
\cos\left(2\phi_{g}\right)\frac{\Gamma^{2}}{4B^{2}+\Gamma^{2}},
\label{eq:cp_asym_final3}
\end{eqnarray}
where we have defined  $\alpha_{2}=\frac{g_{2}^2}{4\pi}$ and 
the flavour projections $K^0_k$
\begin{equation} 
K^0_k= \frac{|Y_{1k}|^2}{{\displaystyle \sum_k}|Y_{1k}|^2}.
\end{equation}

Summing up the contribution from the decays of $\widetilde{N}_{+}$ and
$\widetilde{N}_{-}$, one gets the three contributions to the CP 
asymmetry in Eq.~\eqref{eq:cp_asym_total} 
\begin{eqnarray}
\epsilon_{k}^{S}(T) & = & -K^0_k\frac{|A|}{M}\sin\left(\phi_{A}\right)
\frac{4B\Gamma}{4B^{2}+\Gamma^{2}}\Delta_{BF}(T),
 \nonumber \\
&\equiv& K^0_k \Delta_{BF}(T) \, \bar \epsilon^S ,
\label{eq:cp_asym_2}\\
\epsilon_{k}^{V}(T) 
& = & -\frac{3K^0_{k}\alpha_{2}}{4}\frac{m_2}{M}
\ln\frac{m_2^2}{m_2^2+M^2}\left[\frac{|A|}{M}\sin\left
(\phi_{A}+2\phi_{g}\right)-\frac{B}{M}\sin\left(2\phi_{g}\right)\right]
\Delta_{BF}(T),
\nonumber \\
&\equiv& K^0_k \Delta_{BF}(T) \, \bar \epsilon^V ,
\label{eq:cp_asym_1}
\\
\epsilon_{k}^{I}(T) & = & \frac{3K^0_{k}\alpha_{2}}{2}\frac{m_2}{M}
\frac{|A|}{M}\ln\frac{m_2^2}{m_2^2+M^2}
\sin\left(\phi_{A}\right)
\mbox{cos}\left(2\phi_{g}\right)\frac{\Gamma^{2}}{4B^{2}+\Gamma^{2}}
\Delta_{BF}(T), \nonumber \\
&\equiv& K^0_k \Delta_{BF}(T) \,\bar \epsilon^I ,
\label{eq:cp_asym_3}
\end{eqnarray}
where
\begin{eqnarray}
\Delta_{BF}(T) & = & 
\frac{c^{s}(T) - c^{f}(T)}{c^{s}(T) + c^{f}(T)} .
\end{eqnarray}

The asymmetry in Eq.\eqref{eq:cp_asym_2} is the contribution to the 
lepton asymmetry due to CP violation in RH sneutrino mixing 
discussed in Refs.\cite{soft1,soft2,oursoft}. 
Eqs.\eqref{eq:cp_asym_1} and \eqref{eq:cp_asym_3} 
give the contribution to the lepton asymmetry related to CP violation
in decay and in the interference of mixing and decay. They have
similar parametric dependences as the ones derived in Ref.\cite{soft3}.
However  as explicitly shown in the equations, 
the scalar and fermionic CP asymmetries,  
\eqref{eq:cancel}, cancel each
other at zero temperature. Consequently  we find that, up to second
order in the soft supersymmetry-breaking parameters, all contributions
to the lepton asymmetry in the soft supersymmetry scenario require thermal
effects in order to be significant. 

We finish by noticing that 
in this derivation we have neglected thermal corrections to the CP 
asymmetry from the loops, 
i.e., we have computed the imaginary part of the one-loop graphs using 
Cutkosky's cutting rules at $T=0$.

\section{CP Asymmetry in Quantum Mechanics }
\label{sec:QM}
In this section we recompute the asymmetry using a quantum mechanical (QM)
approach, based on an effective (non hermitic) Hamiltonian
\cite{soft1,soft2,soft3}. In this language an analogy can be drawn between
the $\widetilde N$--$\widetilde N^{\dagger}$ system and the system of 
neutral mesons such as $K^0$--$\overline K^0$ and its time evolution is  
determined in the non-relativistic limit by the Hamiltonian: 

\begin{eqnarray}
H & = & 
\left(\begin{array}{cc} M & \frac{B}{2}\\
\frac{B}{2} & M \end{array}
\right)
-\frac{i}{2}\left(\begin{array}{cc}
\Gamma & \frac{\Gamma A^{*}}{M}\\
\frac{\Gamma A}{M} &  \Gamma\end{array}\right),
\end{eqnarray}
with $\Gamma$ given in Eq.\eqref{eq:gamma}.

In Refs.\cite{soft1,soft2,soft3} the QM formalism was applied for 
weak initial states $\widetilde N$ and $\widetilde N^{\dagger}$. 
In practice it is possible to use the formalism to study the evolution 
of either initially weak or mass eigenstates. So in order to study the 
dependence of the results on the choice of physical 
initial conditions we will compute the asymmetry in this formalism assuming 
either of the two possibilities for initial states.
So we define the basis: 
\begin{eqnarray}
\tilde{N}_{1} & = & \left(a\tilde{N}+b\tilde{N}^{\dagger}\right),
\nonumber \\
\tilde{N}_{2} & = & e^{i\beta}\left(b\tilde{N}-a\tilde{N}^{\dagger}\right).
\label{eq:arbitrary_basis}
\end{eqnarray}
The mass basis, Eq.\eqref{eq:mass_eigenstates} corresponds to 
$(a,b,\beta)=(\frac{1}{\sqrt{2}},\frac{1}{\sqrt{2}},-\frac{\pi}{2})$.
Assuming that the physical initial states were 
pure $\widetilde N$ and $\widetilde N^{\dagger}$  corresponds to 
$(a,b,\beta)=(1,0, \pi)$.

The decay amplitudes of 
$\widetilde N_1$ and $\widetilde N_2$ 
into  fermions   
$f_k=\ell_k^c \tilde h^d$  including the one-loop
contribution from gaugino exchange are:
\begin{eqnarray}
A_{1}^{f_k} & = 
& \left\{Y_{1k} b 
- \frac{3Y_{1k}}{2M^2}\left(a M+bA^{*}\right)
\left(\tilde g_{2}^{*}\right)^{2}\frac{m_2}{16\pi}I_{f} \right\}
[\overline{u}\left(p_{\ell}\right)P_{R}v\left(p_{h} \right)] 
\epsilon_{cd},\nonumber \\
\overline{A_{1}^{\bar f_k}} & = & \left\{Y_{1k} a
- \frac{3Y_{1k}}{2M^2} \left(bM+a A\right)
\left(\tilde g_{2}\right)^{2}\frac{m_2}{16\pi}I_{f} \right\}
[\overline{u}\left(p_{h}\right)P_{L}v\left(p_{\ell}\right)]\epsilon_{cd}, 
\nonumber \\
A_{2}^{f_k} & = & -e^{-i\beta} \left\{Y_{1k} a
- \frac{3Y_{1k}}{2 M^2}
\left(bM-aA^{*}\right)\left(\tilde g_{2}^{*}\right)^{2}\frac{m_2}{16\pi}I_{f}
\right\}
[\overline{u}\left(p_{\ell}\right)P_{R}v\left(p_{h}\right]\epsilon_{cd}
,\nonumber \\
\overline{A _{2}^{\bar f_k}} & = & e^{-i\beta} \left\{ Y_{1k}b
- \frac{3Y_{1k}}{2 M^2}
\left(bA-aM\right)\left(\tilde g_{2}\right)^{2}\frac{m_2}{16\pi}I_{f}
\right\}
[\overline{u}\left(p_{h}\right)P_{L}v\left(p_{\ell}\right)]\epsilon_{cd},
\label{eq:qmamplif}
\end{eqnarray}
where the $\overline{A}$ denotes the decay amplitudes into antifermions.
The corresponding decay amplitudes 
into scalar   $s_k=\tilde{\ell}^c_k h^d$ and 
\begin{eqnarray}
A_{1}^{s_k} & = &\left\{Y_{1k}
\left(a M+bA^{*}\right) 
-\frac{3Y_{1k}}{2}b\left(\tilde g_{2}\right)^{2}
\frac{m_2}{16\pi}I_{s} \right\}\epsilon_{cd}
,\nonumber \\
\overline{A_{1}^{\bar s_k}} & = & \left\{Y_{1k}
\left(bM+a A\right)
-\frac{3Y_{1k}}{2}
a\left(\tilde g_{2}^{*}\right)^{2}\frac{m_2}{16\pi}I_{s}
\right\}\epsilon_{cd}
,\nonumber \\
A_{2}^{s_k} 
& = & e^{-i\beta}\left\{Y_{1k}
\left(bM-aA^{*}\right)
+\frac{3Y_{1k}}{2} a\left(\tilde g_{2}\right)^{2}\frac{m_2}{16\pi}
I_{s} \right\}\epsilon_{cd},\nonumber \\
\overline{A_{2}^{\bar s_k}} & = & e^{-i\beta}\left\{Y_{1k}
\left(bA-aM\right)
-\frac{3Y_{1k}}{2}
b\left(\tilde g_{2}^{*}\right)^{2}\frac{m_2}{16\pi}I_{s}\right\} \epsilon_{cd},
\label{eq:qmamplis}
\end{eqnarray}
where
\begin{eqnarray}
&&\mbox{Re}(I_{f})\equiv f_R  = -\frac{1}{\pi}
\left[\frac{1}{2}\left(\ln\frac{m_{2}^{2}}{m_{2}^{2}+M^{2}}
\right)^{2}+\mbox{Li}_{2}\left(\frac{m_{2}^{2}}{m_{2}^{2}+M^{2}}
\right)-\zeta(2)\right], 
\nonumber  \\
&& \mbox{Re}(I_s) \equiv s_R =  
\frac{1}{\pi}\left[\frac{1}{2} \left(\ln\frac{m_2^{2}}
{m_2^{2}+M^{2}}\right)^{2}+\mbox{Li}_{2}\left(\frac{m_2^{2}}{m_2^{2}+M^{2}}\right)-\zeta(2)+B_{0}\left(M^{2},m_2,0\right)+B_{0}\left(M^{2},0,m_2\right)
\right],
\nonumber 
\\
&&\mbox{Im}(I_f)\equiv f_I=\mbox{Im}(I_s)\equiv s_I=
-\ln\frac{m_2^{2}}{m_2^{2}+M^{2}}.
\end{eqnarray}

The eigenvectors of the Hamiltonian in terms of the 
states $\widetilde N_1$ and $\widetilde N_2$ are:
\begin{eqnarray}
\left|\tilde{N}_{L}\right\rangle  & = 
& \left(ap+bq\right)\left|\tilde{N}_{1}
\right\rangle +e^{-i\beta}\left(bp-aq\right)\left|\tilde{N}_{2}
\right\rangle ,\nonumber \\
\left|\tilde{N}_{H}\right\rangle  & = & 
\left(a p-bq\right)\left|\tilde{N}_{1}\right\rangle 
+e^{-i\beta}\left(bp+aq\right)\left|\tilde{N}_{2}\right\rangle ,
\label{eq:LH_to_arbitrary}
\end{eqnarray}
where
\begin{eqnarray}
\frac{q}{p} & = & -1-\frac{\Gamma|A|}{BM}\sin\left(\phi_{A}\right)
-\frac{\Gamma^{2}|A|^{2}}{M^{2}B^{2}}\cos^{2}\left(\phi_{A}\right)
-\frac{i}{2}\frac{\Gamma^{2}|A|^{2}}{M^{2}B^{2}}\sin\left(2\phi_{A}\right).
\label{eq:ratio_qp}
\end{eqnarray}
At the time $t$ the states $\widetilde N_1$ and $\widetilde N_2$ 
have evolved into 
\begin{eqnarray}
\left|\tilde{N}_{1,2}(t)\right\rangle  
 & = & \frac{1}{2}\left\{ \left[e_{L}(t)+e_{H}(t) \pm C_{0}
\left(e_{L}(t)-e_{H}(t)\right)\right]\left|\tilde{N}_{1,2}\right
\rangle \right.  \nonumber \\
&& \left.+e^{\mp i\left(\beta\right)}C_{1,2}\left(e_{L}(t)
-e_{H}(t)\right)\left|\tilde{N}_{2,1}\right\rangle \right\} \; , 
\label{eq:arbitrary_time_evolution}
\end{eqnarray}
where
\begin{eqnarray}
C_{0}  = a b \left(\frac{p}{q}+\frac{q}{p}\right),
\;\;\;\;
C_{1} = b^{2}\frac{p}{q}-a^{2}\frac{q}{p}, \;\;\;\;
C_{2} &=& b^{2}\frac{q}{p}-a^{2}\frac{p}{q}, \;\;\;\;
\label{eq:C_coeff}
\end{eqnarray}
and 
\begin{eqnarray}
e_{H,L}(t) & \equiv & e^{-i(M_{H,L}-\frac{i}{2}\Gamma_{H,L})t}.
\end{eqnarray}

The total time integrated lepton asymmetry is
\begin{equation}
\epsilon_k^{QM} = \frac{\displaystyle \sum_{i=1,2,a_k} 
\Gamma(\widetilde{N}_i \rightarrow a_k)
- \Gamma(\widetilde{N}_i \rightarrow \bar{a}_k)}
{\displaystyle \sum_{i=1,2,a_k,k} \Gamma(\widetilde{N}_i \rightarrow a_k)
+ \Gamma(\widetilde{N}_i \rightarrow \bar{a}_k)} \ , 
\label{eq:qmcp_asym_total} 
\end{equation}
where $\Gamma(\widetilde{N}_i \rightarrow a_k)$ 
are the time integrated  decay rates which 
from Eq.\eqref{eq:arbitrary_time_evolution} are found to be   
\begin{eqnarray}
\Gamma(\widetilde{N}_i \rightarrow a_k) = \frac{1}{4}
 \frac{c^{a_k}}{ 16\pi M}
&\Big(&\left|A^{a_k}_{i}\right|^{2}G_{i+}+
\left|A_{j\neq i}^{a_k}\right|^{2}G_{j-}
\nonumber \\
&&+2\left[\mbox{Re}\left({A^{a_k}_{i}}^{*}A^{a_k}_{j\neq i}
\right)G_{ii}^{R} -\mbox{Im}\left({A^{a_k}_{i}}^{*}A^{a_k}_{j\neq i}
\right)G_{ii}^{I}\right]\Big).
\label{eq:intdecrat}
\end{eqnarray}
$\Gamma(\widetilde{N}_i \rightarrow \bar a_k)$ can be obtained from
Eq.\eqref{eq:intdecrat} with the replacement 
$A^{a_k}_{l}\rightarrow  \overline{A^{\bar a_k}_{l}}$.
We have defined the time integrated  projections 
\begin{eqnarray}
G_{1,2+}  & = & 
2\left(\frac{1}{1-y^{2}}+\frac{1}{1+x^{2}}\right)+ 2
\left|C_{0}\right|^{2}\left(\frac{1}{1-y^{2}}
-\frac{1}{1+x^{2}}\right)\nonumber \\
 &  & \pm 8\left[\mbox{Re}
\left(C_{0}\right)\frac{y}{1-y^{2}}
-\mbox{Im}\left(C_{0}\right)\frac{x}{1+x^{2}}\right],\label{eq:G_plus}\\
G_{1,2-}  & = & 2\left|C_{1,2}\right|^{2}
\left(\frac{1}{1-y^{2}}-\frac{1}{1+x^{2}}\right),
\nonumber \\
G_{11(22)}^{R} 
 & = & 2\left\{ \mbox{Re}
\left[e^{\mp i\beta}C_{1(2)}\right]
\frac{y}{1-y^{2}}-\mbox{Im}\left[e^{\mp i\beta}
C_{1(2)}\right]\frac{x}{1+x^{2}}\right\} \nonumber \\
 &  & \pm 2
\mbox{Re}\left[e^{\mp i\beta}
C_{0}^{*}C_{1(2)}\right]\left(\frac{1}{1-y^{2}}
-\frac{1}{1+x^{2}}\right),\label{eq:G_real}\\
G_{11(22)}^{I} 
 & = & 2\left\{ 
\mbox{Im}\left[e^{\mp i\beta}C_{1(2)}\right]
\frac{y}{1-y^{2}}+\mbox{Re}\left[e^{\mp i\beta}C_{1(2)}
\right]\frac{x}{1+x^{2}}\right\} \nonumber \\
 &  & \pm 2
\mbox{Im}\left[e^{\mp i\beta}C_{0}^{*}C_{1(2)}\right]
\left(\frac{1}{1-y^{2}}-\frac{1}{1+x^{2}}\right),
\label{eq:G_factors}
\end{eqnarray}
in terms of masses and width differences coefficients \footnote{
We use the expression of $\Gamma_H-\Gamma_L$ from Ref.\cite{soft3}.
Notice that with this definition  $\Gamma_H-\Gamma_L\neq \Gamma_+-\Gamma_-$.}:
\begin{eqnarray}
x & = & \frac{M_H-M_L}{\Gamma}=\frac{B}{\Gamma}
-\frac{1}{2}\frac{\Gamma|A|^{2}}{BM^{2}}\sin^{2}\left(\phi_{A}\right),\\
y & = & \frac{\Gamma_H-\Gamma_L}{2\Gamma}=\frac{|A|}{M}
\cos\left(\phi_{A}\right)-\frac{B}{2M}.
\end{eqnarray}
Substituting Eqs.\eqref{eq:intdecrat}--\eqref{eq:G_factors} one can 
write the numerator in Eq.\eqref{eq:qmcp_asym_total} as
\begin{equation}
\sum_i \Gamma(\widetilde{N}_i \rightarrow a_k)
- \Gamma(\widetilde{N}_i \rightarrow \bar{a}_k)\equiv
\Delta\Gamma^{a_k,R}
+\Delta\Gamma^{a_k,NR}+\Delta\Gamma^{a_k,I},
\end{equation}
with 
\begin{eqnarray}
\Delta\Gamma^{a_k,R}&=& \frac{1}{2}
\frac{c^{a_k}}{16 \pi M} 
\frac{x^{2}+y^{2}}{\left(1-y^{2}\right)\left(1+x^{2}\right)}\Big\{
\left|C_{0}\right|^{2}
\left(\left|A^{a_k}_{1}\right|^{2}
-\left|\overline{A^{\bar a_k}_{1}}\right|^{2}
+\left|A^{a_k}_{2}\right|^{2}
-\left|\overline{A^{\bar a_k}_{2}}\right|^{2}
\right)
\nonumber \\
&&-
\frac{\left(\left|C_{1}\right|^{2}-\left|C_{2}\right|^{2}\right)}{2}
\left(\left|A^{a_k}_{1}\right|^{2}
-\left|\overline{A^{\bar a_k}_{1}}\right|^{2}
-\left|A^{a_k}_{2}\right|^{2}
+\left|\overline{A^{\bar a_k}_{2}}\right|^{2}
\right)
\label{eq:delgm}  \\
&&+2 \left[
\mbox{Re}\left({A^{a_k}_{1}}^{*}A^{a_k}_{2}
-\overline{A^{\bar a_k}_{1}}^{*}\overline{A^{\bar a_k}_{2}}\right)
\mbox{Re}\left(e^{-i\beta}C_{0}^{*}C_{1}\right)
-
\mbox{Re}\left({A^{a_k}_{2}}^{*}A^{a_k}_{1}
+\overline{A^{\bar a_k}_{2}}^{*}\overline{A^{\bar a_k}_{1}}\right)
\mbox{Re}\left(e^{i\beta}C_{0}^{*}C_{2}\right)
\right]
\nonumber \\
&&-2 \left[
\mbox{Im}\left({A^{a_k}_{1}}^{*}A^{a_k}_{2}
-\overline{A^{\bar a_k}_{1}}^{*}\overline{A^{\bar a_k}_{2}}\right)
\mbox{Im}\left(e^{-i\beta}C_{0}^{*}C_{1}\right)
-
\mbox{Im}\left({A^{a_k}_{2}}^{*}A^{a_k}_{1}
+\overline{A^{\bar a_k}_{2}}^{*}\overline{A^{\bar a_k}_{1}}\right)
\mbox{Im}\left(e^{i\beta}C_{0}^{*}C_{2}
\right)\right] \Big\},
\nonumber 
\end{eqnarray}
\begin{eqnarray}
\Delta\Gamma^{a_k,NR}&=&
\frac{c^{a_k}}{16 \pi M} 
\frac{1}{\left(1-y^{2}\right)}\Big\{
2 y \mbox{Re}(C_0)
\left(\left|A^{a_k}_{1}\right|^{2}
-\left|\overline{A^{\bar a_k}_{1}}\right|^{2}
-\left|A^{a_k}_{2}\right|^{2}
+\left|\overline{A^{\bar a_k}_{2}}\right|^{2}
\right)
\nonumber \\
&&+ \left(\left|A^{a_k}_{1}\right|^{2}
-\left|\overline{A^{\bar a_k}_{1}}\right|^{2}
+\left|A^{a_k}_{2}\right|^{2}
-\left|\overline{A^{\bar a_k}_{2}}\right|^{2}
\right)
\label{eq:delgd}  \\
&&+ y\left[
\mbox{Re}\left({A^{a_k}_{1}}^{*}A^{a_k}_{2}
-\overline{A^{\bar a_k}_{1}}^{*}\overline{A^{\bar a_k}_{2}}\right)
\mbox{Re}\left(e^{-i\beta}C_{1}\right)
+
\mbox{Re}\left({A^{a_k}_{2}}^{*}A^{a_k}_{1}
-\overline{A^{\bar a_k}_{2}}^{*}\overline{A^{\bar a_k}_{1}}\right)
\mbox{Re}\left(e^{i\beta}C_{2}\right)
\right]
\nonumber \\
&&-y \left[
\mbox{Im}\left({A^{a_k}_{1}}^{*}A^{a_k}_{2}
-\overline{A^{\bar a_k}_{1}}^{*}\overline{A^{\bar a_k}_{2}}\right)
\mbox{Im}\left(e^{-i\beta}C_{1}\right)
+
\mbox{Im}\left({A^{a_k}_{2}}^{*}A^{a_k}_{1}
-\overline{A^{\bar a_k}_{2}}^{*}\overline{A^{\bar a_k}_{1}}\right)
\mbox{Im}\left(e^{i\beta}C_{2}
\right)\right]\Big\}, \nonumber
\end{eqnarray}
\begin{eqnarray}
\Delta\Gamma^{a_k,I}&=&
\frac{c^{a_k}}{16 \pi M} 
\frac{x}{\left(1+x^{2}\right)}\Big\{
-2  \mbox{Im}(C_0)
\left(\left|A^{a_k}_{1}\right|^{2}
-\left|\overline{A^{\bar a_k}_{1}}\right|^{2}
-\left|A^{a_k}_{2}\right|^{2}
+\left|\overline{A^{\bar a_k}_{2}}\right|^{2}
\right) \label{eq:delgi}  \\ 
&&- \left[
\mbox{Re}\left({A^{a_k}_{1}}^{*}A^{a_k}_{2}
-\overline{A^{\bar a_k}_{1}}^{*}\overline{A^{\bar a_k}_{2}}\right)
\mbox{Re}\left(e^{-i\beta}C_{1}\right)
+
\mbox{Re}\left({A^{a_k}_{2}}^{*}A^{a_k}_{1}
-\overline{A^{\bar a_k}_{2}}^{*}\overline{A^{\bar a_k}_{1}}\right)
\mbox{Re}\left(e^{i\beta}C_{2}\right)
\right]
\nonumber \\
&&- \left[
\mbox{Im}\left({A^{a_k}_{1}}^{*}A^{a_k}_{2}
-\overline{A^{\bar a_k}_{1}}^{*}\overline{A^{\bar a_k}_{2}}\right)
\mbox{Im}\left(e^{-i\beta}C_{1}\right)
+
\mbox{Im}\left({A^{a_k}_{2}}^{*}A^{a_k}_{1}
-\overline{A^{\bar a_k}_{2}}^{*}\overline{A^{\bar a_k}_{1}}\right)
\mbox{Im}\left(e^{i\beta}C_{2}
\right)\right]\Big\} \nonumber \;. 
\end{eqnarray}
In writing the above equations we have classified  the contributions 
as {\sl resonant},  ({\sl non-resonant}), $R$ 
($NR$), depending on whether
they present an overall factor $\frac{x^2+y^2}{1+x^2}$
(or no $\frac{1}{1+x^2}$ at all).
We have labeled the remainder as {\sl interference} term, $I$. 
 
After substituting the explicit values for the amplitudes and the
coefficients and neglecting all those terms which cancel in both
basis we get that 
\begin{eqnarray}
\Delta\Gamma^{f_k,R}=-c^{f} \Delta\Gamma_k^{R},
&&\;\;\;\;\;\;
\Delta\Gamma^{s_k,R}=c^{s} \Delta\Gamma_k^{R}, \nonumber \\
\Delta\Gamma^{f_k,NR}=-c^{f} \Delta\Gamma_k^{NR},
&&\;\;\;\;\;\;
\Delta\Gamma^{s_k,NR}=c^{s} \Delta\Gamma_k^{NR}, \nonumber \\
\Delta\Gamma^{f_k,I}=-c^{f} \Delta\Gamma_k^{I},
&&\;\;\;\;\;\;
\Delta\Gamma^{s_k,I}=c^{s} \Delta\Gamma_k^{I},  \label{eq:qmcancel}
\end{eqnarray}
with
\begin{eqnarray}  
\Delta\Gamma_k^{R}
&=& -\frac{1}{4\pi}Y_{1k}^2 \left[ (a^2-b^2)^2+ (2ab)^2 \cos 2\beta\right]
|A| \sin(\phi_A) \frac{1}{x} \frac{x^2+y^2}{(1-y^2)(1+x^2)}, \\
\Delta\Gamma_k^{NR}&=&
\frac{3}{16\pi} Y_{1k}^2 \alpha_2  \ln\frac{m_2^2}{m_2^2+M^2} 
\frac{m_2}{M}  \frac{1}{1-y^2}
\left[- |A| \sin(\phi_A+2\phi_g) \right.\nonumber \\
     && \;\; \left. + y M \left(2 (2ab)^2+ (a^2-b^2)^2\cos(2\beta)\right)
\sin(2\phi_g)\right], \\
\Delta\Gamma_k^{I}&=&
\frac{3}{16\pi} Y_{1k}^2 \alpha_2 \ln\frac{m_2^2}{m_2^2+M^2} 
\frac{m_2}{M}
\frac{1}{1+x^2} |A|\sin(\phi_A)\cos(2\beta)\cos(2\phi_g). 
\end{eqnarray}      

Eq.\eqref{eq:qmcancel} explicitly displays the cancellation of the 
asymmetries at $T=0$ also in this formalism for either initial 
mass or weak eigenstate right-handed sneutrinos. \footnote{
We have traced the discrepancy with Ref.\cite{soft3} to a missing
$\cos(\phi_f-\phi_s)$ factor in their expression for $\sin\delta_s$ 
in their Eq.(19).
Once that factor is included,  $\epsilon_2^m$ in their Eq.(22) cancels
against  $\epsilon_2^{mdi}$ in their Eq.(25), and  $\epsilon^i$ in  
their Eq.(23) cancels against  $\epsilon^d$ in their Eq.(24) so that
the total asymmetry is zero at $T=0$.}

Introducing the explicit values for the coefficients for 
initial weak RH sneutrinos and the expressions for $x$ and $y$ and
expanding at order $\delta_S^2$ we get
\begin{eqnarray} 
\epsilon_{k}^{R,QMw}(T) & = & -K^0_k\frac{|A|}{M}\sin\left(\phi_{A}\right)
\frac{B\Gamma}{B^{2}+\Gamma^{2}}\Delta_{BF}(T),
\label{eq:qmw_asym_2}\\
\epsilon_{k}^{NR,QMw}(T) 
& = & -\frac{3K^0_{k}\alpha_{2}}{4}\frac{m_2}{M}
\ln\frac{m_2^2}{m_2^2+M^2}\Big[\frac{|A|}{M}\sin(\phi_{A})
\cos(2\phi_{g})
\nonumber \\ && 
\;\;\;\;\;\;\;\;\;\;\;\;\;\;\;\;\;\;\;\;\;\;\;
\;\;\;\;\;\;\;\;\;\;\;\;\;\;\;\;\;\;\;\;\;\;\;
+\frac{B}{2M}\sin\left(2\phi_{g}\right)\Big]
\Delta_{BF}(T),
\label{eq:qmw_asym_1}
\\
\epsilon_{k}^{I,QMw}(T) & = & \frac{3K^0_{k}\alpha_{2}}{4}\frac{m_2}{M}
\frac{|A|}{M}\ln\frac{m_2^2}{m_2^2+M^2}
\sin\left(\phi_{A}\right)
\mbox{cos}\left(2\phi_{g}\right)\frac{\Gamma^{2}}{B^{2}+\Gamma^{2}} 
\Delta_{BF}(T)\;.
\label{eq:qmw_asym_3}
\end{eqnarray}
Correspondingly for initial $\widetilde N_\pm$ states one gets 
\begin{eqnarray} 
\epsilon_{k}^{R,QMm}(T) & = & K^0_k\frac{|A|}{M}\sin\left(\phi_{A}\right)
\frac{B\Gamma}{B^{2}+\Gamma^{2}}\Delta_{BF}(T),
\label{eq:qmm_asym_2}\\
\epsilon_{k}^{NR,QMm}(T) 
& = & -\frac{3K^0_{k}\alpha_{2}}{4}\frac{m_2}{M}
\ln\frac{m_2^2}{m_2^2+M^2}\Big[\frac{|A|}{M}\sin(\phi_{A})
\cos(2\phi_{g})
\nonumber \\ && \;\;\;\;\;\;\;\;\;\;\;\;\;\;\;\;\;\;\;\;\;\;\;
\;\;\;\;\;\;\;\;\;\;\;\;\;\;\;\;\;\;\;\;\;\;\;
+\frac{B}{2M}\sin\left(2\phi_{g}\right)\Big]
\Delta_{BF}(T),
\label{eq:qmm_asym_1}
\\
\epsilon_{k}^{I,QMm}(T) & = & -\frac{3K^0_{k}\alpha_{2}}{4}\frac{m_2}{M}
\frac{|A|}{M}\ln\frac{m_2^2}{m_2^2+M^2}
\sin\left(\phi_{A}\right)
\mbox{cos}\left(2\phi_{g}\right)\frac{\Gamma^{2}}{B^{2}+\Gamma^{2}} 
\Delta_{BF}(T) \;.
\label{eq:qmm_asym_3}
\end{eqnarray}
Comparing Eqs.\eqref{eq:qmm_asym_2}--\eqref{eq:qmm_asym_3}
with Eqs.\eqref{eq:qmw_asym_2}--\eqref{eq:qmw_asym_3}
and Eqs.\eqref{eq:cp_asym_2}--\eqref{eq:cp_asym_3}
we find that they show very similar parametric dependence though there are 
some differences in the numerical coefficients. 
In particular we find that $\epsilon_k^{R,QM}$, $\epsilon_k^{I,QM}$ 
and the $B$-dependent (second term) 
in either the weak or mass basis $\epsilon_k^{NR,QM}$ 
coincide with  $\epsilon_k^{S}$, 
$\epsilon_k^{I}$ and the 
B-dependent term in $\epsilon_k^{V}$ derived in the
previous section with the redefinition 
$A\rightarrow 2 A$, $B\rightarrow 2 B$ and 
$\sin(\phi_A)\rightarrow\pm \sin(\phi_A)$. We find only  
some differences in the phase combination which appears in the $B$ 
independent term in the asymmetries  $\epsilon_k^{I,QM}$ and
$\epsilon_k^{V}$ as seen in 
Eqs.\eqref{eq:cp_asym_1} and Eqs.\eqref{eq:qmw_asym_1} 
and \eqref{eq:qmm_asym_1}.

\section{Results}
\label{sec:results}
Next we quantify the parameters for which successful leptogenesis
induced by the different sources of CP violation discussed 
in the previous sections is possible
by solving the corresponding set of Boltzmann Equations (BE).

The relevant classical BE describing the decay, inverse 
decay and scattering processes involving the sneutrino states
in the framework of flavoured soft leptogenesis were derived 
in detail in Ref.\cite{oursoft} assuming that the physically 
relevant sneutrino states were the mass eigenstates 
\eqref{eq:mass_eigenstates}. As we have seen in the previous
section, the choice of the physical basis 
for the sneutrino states does not lead to important differences
in the parametric form of the generated asymmetries, thus in what 
follows we will present our results assuming that the relevant sneutrino
states are the mass eigenstates.

Including  the $\widetilde{N}_{\pm}$ and $N$ decay and 
inverse decay processes as well as all the $\Delta L=1$ scattering 
processes induced by the $top$ Yukawa coupling 
the final set of BE takes the form
\footnote{$\Delta L=1$ scattering involving gauge bosons, $\Delta L=2$ 
off-shell scattering processes involving the pole-subtracted s-channel 
and the u and t-channel, as well as the  
the $L$ conserving processes from $N$ and $\widetilde{N}$ pair creation 
and annihilation have not been included. The
reaction rates for these processes are quartic in the Yukawa 
couplings and can  be safely neglected.} :  
\begin{eqnarray}
sHz\frac{dY_{N}}{dz} & = & 
-\left(\frac{Y_{N}}{Y_{N}^{eq}}-1\right)
\left(\gamma_{N}+4\gamma_{t}^{(0)}+4\gamma_{t}^{(1)}
+4\gamma_{t}^{(2)}+2\gamma_{t}^{(3)}+4\gamma_{t}^{(4)}\right),
\label{eq:BEN}\\
sHz\frac{dY_{\widetilde{N}_{\mbox{tot}}}}{dz} & = & 
-\left(\frac{Y_{\widetilde{N}_{\mbox{tot}}}}{Y_{\widetilde{N}}^{eq}}
-2\right)\left(\gamma_{\widetilde{N}}+\gamma_{\widetilde{N}}^{(3)}
+3\gamma_{22}+2\gamma_{t}^{(5)}+2\gamma_{t}^{(6)}
+2\gamma_{t}^{(7)}+\gamma_{t}^{(8)}+2\gamma_{t}^{(9)}\right)
\nonumber \\
&&
+\gamma_{\widetilde{N}}\sum_k\epsilon_k(T) \frac{Y_{\Delta_k}}{2Y_{c}^{eq}}, 
\label{eq:BENt} 
\end{eqnarray}
\begin{eqnarray}
sHz\frac{dY_{\Delta_{k}}}{dz} 
& = & -\left\{\epsilon^{k}(T)
\left(\frac{Y_{\widetilde{N}_{\mbox{tot}}}}{Y_{\widetilde{N}}^{eq}}-2\right) 
\gamma_{\widetilde{N}} -\sum_{j}A_{kj}\frac{Y_{\Delta_{j}}} 
{2Y_{c}^{eq}} \, 
\gamma_{\widetilde{N}}^{(k)}\right.
\nonumber \\
 &  & -\sum_{j}A_{kj}
\frac{Y_{\Delta_{j}}}{2Y_{c}^{eq}}
\left(\frac{Y_{\widetilde{N}_{\mbox{tot}}}}{Y_{\widetilde{N}}^{eq}}
\gamma_{t}^{(5)k}+2\gamma_{t}^{(6)k}+2\gamma_{t}^{(7)k}
+\frac{Y_{N}}{Y_{N}^{eq}}\gamma_{t}^{(3)k}+2\gamma_{t}^{(4)k}
\right.\nonumber \\
 &  & \left.\frac{1}{2}\gamma_{N}^{k}+\gamma_{\widetilde{N}}^{(2)k}
+\frac{1}{2}\frac{Y_{\widetilde{N}_{\mbox{tot}}}}{Y_{\widetilde{N}}^{eq}}
\gamma_{t}^{(8)k}+2\gamma_{t}^{(9)k}+2\frac{Y_{N}}{Y_{N}^{eq}}
\gamma_{t}^{(0)k}+2\gamma_{t}^{(1)k}+2\gamma_{t}^{(2)k}
\right)\nonumber \\
 &  & \left.-\sum_{j}A_{kj}\frac{Y_{\Delta_{j}}}
{2Y_{c}^{eq}}\left(2+\frac{1}{2}\frac{Y_{\widetilde{N}_{\mbox{tot}}}}
{Y_{\widetilde{N}}^{eq}}\right)\gamma_{22}^{k} 
\right\}.
\label{eq:BE_L_tot_flavor}
\end{eqnarray}
In the equations above $Y_{c}^{eq}\equiv\frac{15}{4\pi^{2}g_{s}^{*}}$  
and $Y^{\rm eq}_{\tilde N}(T\gg M) = 90 \zeta(3)/(4\pi^4g_s^*)$,    
where $g_{s}^{*}$ is the total number of entropic degrees of 
freedom, $g_{s}^{*}=228.75$ in the MSSM. 

Before proceeding let's point out that, as recently discussed
in Refs.~\cite{riottoqbe}, for
resonant scenarios, the use of quantum BE~\cite{buchqbe,riottoqbe}
(QBE)  may be relevant. For standard resonant leptogenesis
they induce a $T$ dependence  in the CP asymmetry which can enhance 
the produced baryon number. 
However, as shown in Ref.\cite{ourqbe}
for soft leptogenesis from CP violation in mixing, due to the 
thermal nature of the 
mechanism already at the classical level, the introduction of
quantum effects does not lead to such enhancement and 
the required values of the parameters 
to achieve successful leptogenesis are not substantially modified.
Thus in this work we study the impact of CP violation in 
sneutrino decay and the interference of mixing and decay
on the values of the parameters required for successful leptogenesis  
in the context of the classical BE as described above. 

In  writing the BE relevant in the regime in which flavours
have to be considered \cite{flavour1,flavour2,db1,barbieri}, 
it is most appropriate  to follow the evolution of
$Y_{\Delta_{k}}$ where 
$\Delta_{k}
=\frac{B}{3}-Y_{L_{k_{f}}}-Y_{L_{k_{s}}}\equiv\frac{B}{3}
-Y_{L^k_{\rm tot}}$.
This is so because  $\Delta_{k}$ is conserved by sphalerons and by 
other MSSM interactions. 
In particular, notice that the MSSM processes enforce the equality
of fermionic and scalar lepton asymmetries of the same flavour. 

In Eqs.\eqref{eq:BEN}--\eqref{eq:BE_L_tot_flavor}
we have accounted for the CP asymmetries 
in the $\widetilde{N}_{\pm}$  two body decays to the order 
described in the previous sections.  
We have neglected higher order terms in supersymmetry-breaking 
parameters which 
could lead to differences in the distribution and thermal widths of 
$\widetilde  N_\pm$ 
and correspondingly we have written a unique BE for 
$Y_{\widetilde{N}_{\mbox{tot}}}\equiv 
Y_{\widetilde{N}_{+}}+Y_{\widetilde{N}_{-}}$. 

In Eq.~(\ref{eq:BE_L_tot_flavor}) we have defined the flavoured
thermal widths
\begin{eqnarray}
\gamma_{\widetilde{N}}^{k}= K^0_{k}\,\gamma_{\widetilde{N}},
\;\;\;\;\;\;\;\;\;\;
\gamma_{t}^{(l)k}= K^0_k \,\gamma_{t}^{(l)},
\end{eqnarray}
where 
the different $\gamma$'s are the thermal widths for the following processes
(in all cases a sum over the CP conjugate final states is implicit):
\begin{eqnarray}
&&\gamma_{\widetilde N}=\gamma^f_{\widetilde N}+\gamma^s_{\widetilde N}=
\gamma(\widetilde{N}_{\pm}\leftrightarrow
\bar{\tilde{h}}\ell)
+\gamma(\widetilde{N}_{\pm} \leftrightarrow h\tilde{\ell}) ,
\nonumber \\
&&\gamma^{(3)}_{\widetilde N}=\gamma( 
\widetilde{N}_{\pm}\leftrightarrow 
\tilde{\ell}^{*}\tilde{u}\tilde{q})\; , 
\nonumber \\
&&\gamma_{22} =  \gamma(\widetilde{N}_{\pm}
\tilde{\ell}\leftrightarrow\tilde{u}\tilde{q}
)=\gamma(\widetilde{N}_{\pm}
\tilde{q}^{*}\leftrightarrow\tilde{\ell}^{*}\tilde{u}
)=\gamma(\widetilde{N}_{\pm}\tilde{u}^{*}
\leftrightarrow\tilde{\ell}^{*}\tilde{q}) ,
\nonumber \\
&&\gamma_t^{(5)}=\gamma(\widetilde{N}_{\pm}
\ell\leftrightarrow q\tilde{u})=\gamma(
\widetilde{N}_{\pm}\ell\leftrightarrow\tilde{q}\bar{u})\; ,
\nonumber \\
&&\gamma_t^{(6)}=\gamma(
\widetilde{N}_{\pm}\tilde{u}\leftrightarrow\bar{\ell}q)=\gamma( 
\widetilde{N}_{\pm}\tilde{q}^{*}\leftrightarrow\bar{\ell}\bar{u})\;,
\nonumber \\
&&\gamma_t^{(7)}=\gamma(
\widetilde{N}_{\pm}\bar{q}\leftrightarrow\bar{\ell}\tilde{u})=\gamma( 
\widetilde{N}_{\pm}u\leftrightarrow\bar{\ell}\tilde{q}) , 
\nonumber \\
&&\gamma_t^{(8)}=\gamma(
\widetilde{N}_{\pm}\tilde{\ell}^{*}\leftrightarrow\bar{q}u) ,
\nonumber \\
&&\gamma_t^{(9)}=\gamma(
\widetilde{N}_{\pm}q\leftrightarrow\tilde{\ell}u)= 
\gamma(\widetilde{N}_{\pm}\bar{u}\leftrightarrow\tilde{\ell}\bar{q}) ,
\nonumber \\
&&\gamma_N=\gamma(N\leftrightarrow \ell h)+
\gamma(N\leftrightarrow \tilde{\ell}^* \tilde h) ,
\nonumber \\
&&\gamma_t^{(0)}=\gamma(N\tilde{\ell}\leftrightarrow q\tilde{u})=
\gamma(N\tilde{\ell}\leftrightarrow\tilde{q}\bar{u}) ,
\nonumber \\
&&\gamma_t^{(1)}=\gamma(N\bar{q}\leftrightarrow\tilde{\ell}^{*}\tilde{u})=
\gamma(N\leftrightarrow\tilde{\ell}^{*}\tilde{q})\; ,
\nonumber \\
&&\gamma_t^{(2)}=\gamma(N\tilde{u}^{*}\leftrightarrow\tilde{\ell}^{*}q)=
\gamma(N\tilde{q}^{*}\leftrightarrow\tilde{\ell}^{*}\bar{u})\; , 
\nonumber \\
&&\gamma_t^{(3)}=\gamma(N\ell\leftrightarrow q\bar{u})\; ,
\nonumber \\
&&\gamma_t^{(4)}=\gamma (N\leftrightarrow\bar{\ell}q)=
\gamma(N\bar{q}\leftrightarrow\bar{\ell}\bar{u})\; .
\label{eq:gammas}
\end{eqnarray}
The explicit expressions for 
the $\gamma$'s in Eq.~(\ref{eq:gammas}) can be found, for example, 
in \cite{plumacher} for the case of 
Boltzmann-Maxwell distribution functions and neglecting Pauli-blocking 
and stimulated emission as well as the relative motion of the particles
with respect to the plasma.

The value of $A_{\alpha\beta}$ depends on which processes
are in thermal equilibrium when leptogenesis is taking place. 
As we will see below for any of the considered sources of CP violation, 
the relevant temperature window around $M \sim T$ corresponds to 
$T<(1+\tan^{2}\beta)\times 10^{9}\mbox{GeV}$. In this regime
the processes mediated by all the three charged lepton $(e,\mu,\tau)$ Yukawa 
couplings are in equilibrium i.e. they are faster than the processes involving
$\widetilde{N}_{\pm}$ and one gets~\cite{antusch}
\begin{equation}
A=\left(\begin{array}{ccc}
-\frac{93}{110} & \frac{6}{55} & \frac{6}{55}\\
\frac{3}{40} & -\frac{19}{30} & \frac{1}{30}\\
\frac{3}{40} & \frac{1}{30} & -\frac{19}{30}
\end{array}\right).
\label{eq:Athree}
\end{equation}
After conversion by the sphaleron transitions, the final baryon asymmetry
is given by 
\begin{equation}
Y_{B}= \frac{24+4n_H}{66+13n_H}
Y_{B-L}(z\rightarrow \infty)
=\frac{8}{23} \, \sum_k Y_{\Delta_k}(z\rightarrow \infty),
\label{eq:yb}
\end{equation}
where ${\displaystyle \sum_k} Y_{\Delta_k}(z\rightarrow \infty)$ can be parametrized as 
\begin{equation}
\sum_k Y_{\Delta_k}(z\rightarrow \infty) =
-2 (\bar \epsilon^S +\bar \epsilon^V +\bar \epsilon^I) \,  
\eta_{\rm fla} \;,
\label{eq:etadef}
\end{equation}
with $\bar \epsilon^S$, $\bar \epsilon^V$, and $\bar \epsilon^I$
are given in Eqs.\eqref{eq:cp_asym_2}--\eqref{eq:cp_asym_3}.  
$\eta_{\rm fla}$ is the dilution factor which
takes into account the possible inefficiency in the production of the
singlet sneutrinos, the erasure of the generated asymmetry by
$L$-violating scattering processes and the temperature and flavour 
dependence of the CP asymmetry. It is obtained by solving the array
of BE above.  Within our approximations for the 
thermal widths, $\eta_{\rm fla}$ depends on the flavour projections 
$K^0_k$, on  the Yukawa couplings 
$(YY^\dagger)_{11}$ and on the heavy mass $M$, with the dominant dependence 
on these last two arising in the combination
\begin{equation}
(YY^\dagger)_{11}\,  v_u^2 \equiv m_{eff}\, M,
\label{eq:meff}
\end{equation}
where $v_u$ is the vacuum expectation 
value of the up-type Higgs doublet, $v_u=v\, \sin\beta $ ($v$=174 GeV) .
There is a residual dependence on $M$ due to the running of 
the top Yukawa coupling as well as the thermal effects included in 
$\Delta_{BF}$ although it is very mild. Also, 
as long as $\tan\beta$ is not very close to one,
the dominant dependence on $\tan\beta$ arises via $v_u$ as given in
Eq.~(\ref{eq:meff}) and it is therefore very mild. 

\FIGURE[t]{
\includegraphics[width=0.6\textwidth]{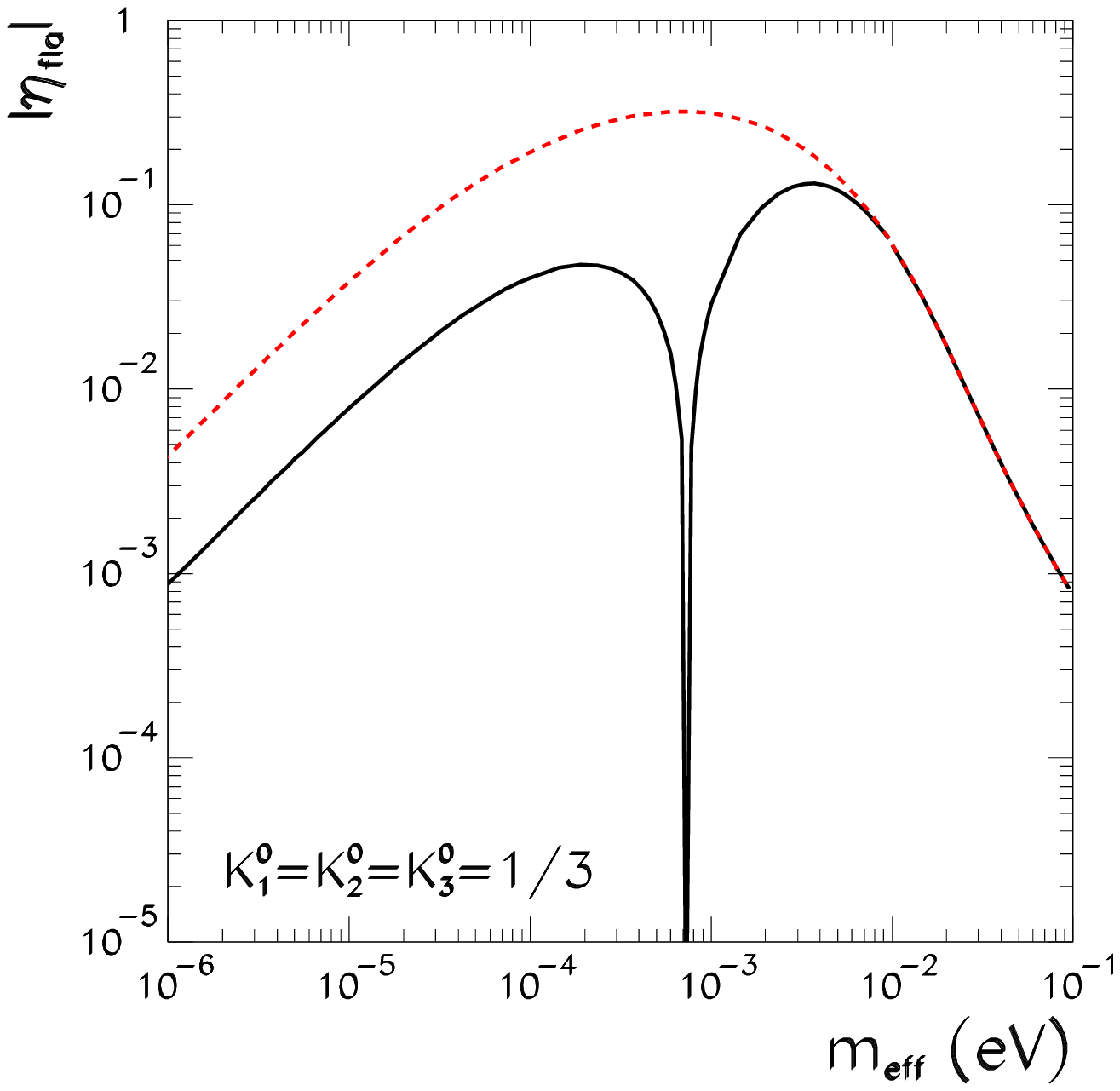}
\caption{Efficiency factor $|\eta_{\rm fla} |$ as a function of $m_{eff}$ 
for $M=10^{7}$ GeV and $\tan\beta=30$ and for 
$K^0_1=K^0_2=K^0_3=1/3$. 
The two curves correspond to  
vanishing initial $\tilde N$ abundance 
(solid black curve) and thermal initial $\tilde N$ abundance, 
(dashed red curve). 
\label{fig:etaunf}}}

In Fig.~\ref{fig:etaunf} we plot $|\eta_{\rm fla}|$ as a function of $m_{eff}$ for
$M=10^7$ GeV and for the the equally distributed flavour composition 
$K^0_1=K^0_2=K^0_3=1/3$. We consider 
two different initial conditions for the sneutrino abundance.
In one case, one assumes that the ${\tilde N}$ population is created
by their Yukawa interactions with the thermal
plasma, and set $Y_{\tilde N}(z\rightarrow 0)=0$. The other case corresponds to
an initial $\tilde N$ abundance equal to the thermal one, 
$Y_{\tilde N}(z\rightarrow 0)= Y_{\tilde N}^{eq}(z\to 0)$.
As discussed in Ref.\cite{oursoft} for zero initial conditions,
$\eta$ can take both signs depending on the value
of $m_{eff}$ while for thermal initial
conditions, on the contrary, $\eta>0$.

\FIGURE[t]{
\includegraphics[width=\textwidth]{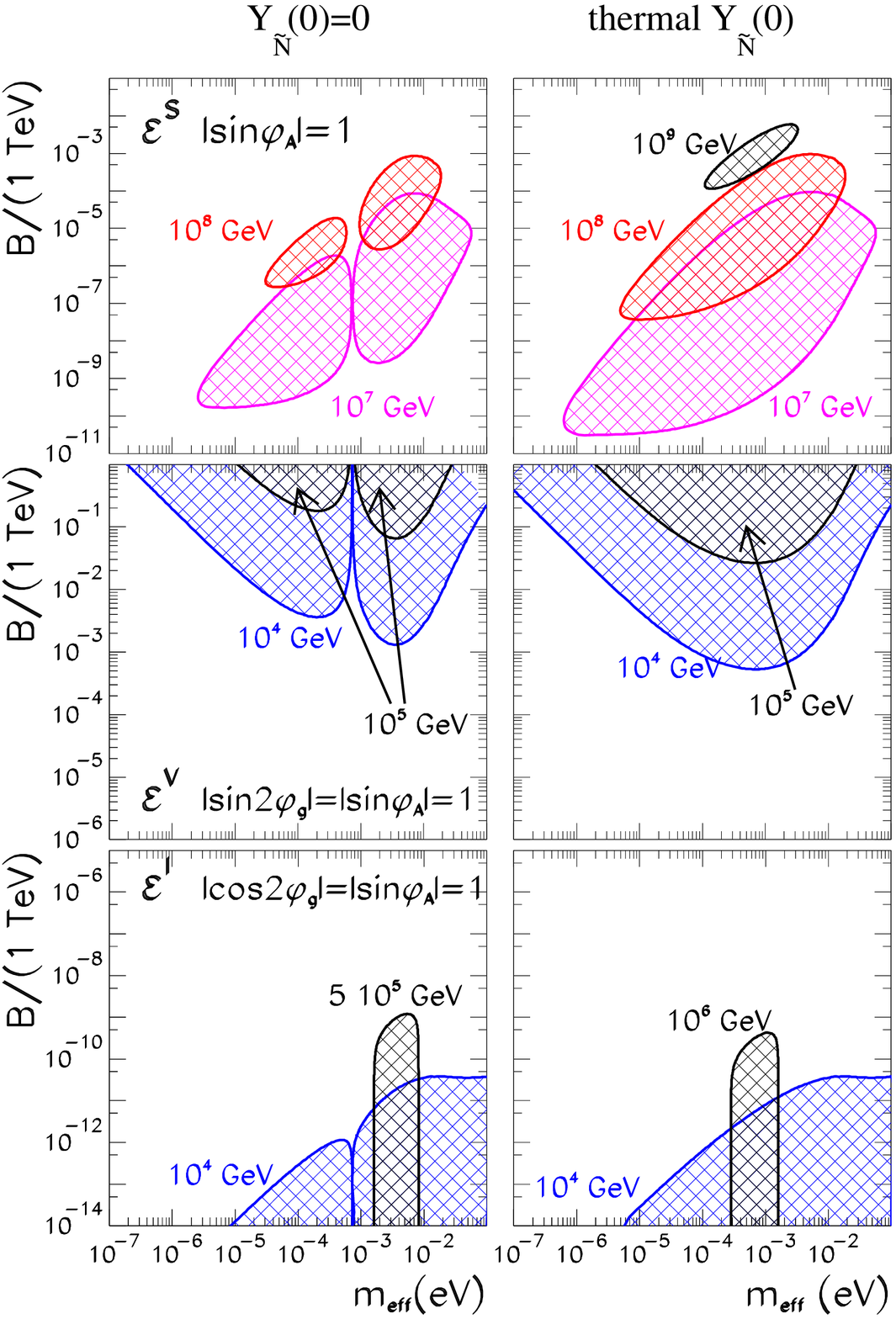}
\caption{
$B, m_{eff}$ regions in which successful soft leptogenesis can
be achieved when flavour effects are included with $K^0_1=K^0_2=K^0_3=1/3$
and for different sources of CP violation.
In all cases we take $A|=m_2=10^3$ GeV and 
$\tan\beta=30$ and different
values of $M$  and $\phi_A$ and $\phi_g$ as labeled in the figure
(see text for details).
The left (right) panels correspond to  vanishing (thermal) 
initial $\tilde N$ abundance .
\label{fig:Bmeffnw}}}
 
Introducing the resulting $\eta_{\rm fla}$ in Eqs.\eqref{eq:etadef} and 
\eqref{eq:yb} we can easily quantify the allowed ranges of 
parameters for which enough asymmetry, 
$Y_B\geq 8.54\times 10^{-11}$ ~\cite{lastwmap},  is
generated.  We plot in  Fig. \ref{fig:Bmeffnw} the resulting 
ranges for  $B$ and $m_{eff}$ for
the equally distributed flavour composition 
$K^0_1=K^0_2=K^0_3=1/3$ and for $|A|=m_2=1$ TeV and 
$\tan\beta=30$. For different values of the CP phases and 
$M$ as explicitly given in the figure. 
 
The upper panels in  Fig. \ref{fig:Bmeffnw} give the 
parameters regions  for which the  CP violation from pure mixing 
effects, $\epsilon^S$, 
can produce the observed asymmetry 
as previously described in 
\cite{soft1,soft2,oursoft}. Due to the resonant nature 
of this contribution, these effects are only large enough for
$B\sim {\cal O}(\Gamma)$ which leads to the well-known condition 
of the unconventionally small values of $B$ and to the
upper bound $M\lesssim 10^{9}$ GeV.

The central panels of Fig.\ref{fig:Bmeffnw} 
give the corresponding  regions for which  
CP violation from gaugino-induced vertex effects, 
$\epsilon^V$, can 
produce the observed baryon asymmetry. Despite being higher order  
in $\delta_S$ and including a  loop suppression 
factor, $\alpha_2$, this contribution can be relevant because it
is dominant for conventional values of the $B$ parameter.
However, in order to overcome the loop and 
$\delta_S$  suppressions this contribution can only be sizeable 
for lighter values of the RH sneutrino masses 
$M\lesssim \;10^{6}$ GeV (within the approximation used
in this work: $\delta_S\ll 1$, $|A|, m_2\sim {\cal O}({\rm TeV})$). 

The parameters chosen in the
figure are such that the second term in Eq.\eqref{eq:cp_asym_1} 
dominates so that the allowed region depicts a lower bound on $B$.
Conversely,  when the first term in Eq.\eqref{eq:cp_asym_1} dominates,
$\epsilon^V$ 
becomes independent of $B$. In this case,
for a given value of $M$ and $\delta_S$  the produced baryon asymmetry can be 
sizeable within the range of $m_{eff}$ values  for which 
$\eta_{\rm fla}$ is large enough. For example for $M= 10^{5}$ GeV,
and $m_2=|A|=1$ TeV and $|\sin(\phi_A+2\phi_g)=1|$ with vanishing 
initial conditions  
\begin{equation} 
10^{-5}<\frac{m_{eff}}{\rm eV}<6.5\times 10^{-4} 
\;\;\;\;\;\;\; {\rm or}\;\;\;\;\;\;
8\times 10^{-4}<\frac{m_{eff}}{\rm eV}<3\times 10^{-2}, 
\end{equation}
where each range corresponds to a sign of the CP phase $\sin(\phi_A+2\phi_g)$ 

Finally we show in the lower panels of Fig.\ref{fig:Bmeffnw} 
the values of $B$ and $m_{eff}$ for which enough baryon asymmetry 
can be generated from the interference of mixing and vertex corrections
$\epsilon^I$, Eq.\eqref{eq:cp_asym_3}. Generically $\epsilon^I$
, is  subdominant  to $\epsilon^S$
since both involve the same CP phase
$\sin(\phi_A)$ while $\epsilon^I$ 
has additional $\delta_S$ and 
loop suppressions:
\begin{equation}
\frac{\bar\epsilon^I}{\bar\epsilon^S}=\frac{-3}{8} \alpha_2 
\frac{m_2}{M}  \ln\frac{m_{2}^{2}}{M^{2}+m_2^{2}} 
\cos(2\phi_g) \frac{\Gamma}{B} \; . 
\end{equation}
Consequently as seen in the above equation  and illustrated in the figure,  
$\epsilon^I$ can only dominate for extremely low values of 
$B$  ($B\ll \Gamma$) for 
which it becomes independent of $B$. Also  we notice
that for $M\lesssim 10^4$ GeV  and $m_{eff}\gtrsim 10^{-2}$ eV
the resulting baryon asymmetry generated 
by this contribution becomes independent of $m_{eff}$ since  
the $m_{eff}^2$ 
dependence from $\Gamma^2$ cancels the approximate 
$1/m_{eff}^2$ dependence of $\eta_{\rm fla}$ in this strong washout regime.

In summary in this work we have quantified in detail the 
contributions to CP violation in right-handed sneutrino decays 
induced by soft supersymmetry-breaking gaugino masses
paying special attention to the role of thermal effects. 
Using a field-theoretical as well as a quantum mechanical 
approach we conclude that for all the soft supersymmetry-breaking 
sources of CP violation considered,   
an exact cancellation between the asymmetries produced 
in the fermionic and bosonic channels occurs 
at  $T=0$ up to second order in 
soft supersymmetry-breaking parameters.  
However, once thermal effects are included the new sources of
CP violation induced by soft supersymmetry-breaking gaugino masses
can be sizeable and they can produce the observed baryon asymmetry
for conventional values of the $B$ parameter.

\acknowledgments 
We thank Y. Nir, Y. Grossman  and S. Davidson 
for discussions and comments. We are specially indebted to
N. Rius for her valuable help in many stages of this work.
This work is supported by National Science Foundation
grant PHY-0354776 and  by Spanish Grants 
FPA-2007-66665-C02-01 and FPA2006-28443-E.

\end{document}